\documentclass[aps,prl,twocolumn,showpacs,superscriptaddress,floatfix]{revtex4}

\usepackage[dvips]{graphicx}
\usepackage{xspace}
\usepackage{latexsym}
\usepackage{amssymb}

\begin{document}

\title{Evidence for an oscillatory signature in atmospheric neutrino 
oscillation}

\date{\today}

\newcommand{\icrr}{\affiliation{Kamioka Observatory, Institute for Cosmic Ray Research, University of Tokyo, Kamioka, Gifu, 506-1205, Japan}}
\newcommand{\ncen}{\affiliation{Research Center for Cosmic Neutrinos, Institute for Cosmic Ray Research, University of Tokyo, Kashiwa, Chiba 277-8582, Japan}}
\newcommand{\bu}{\affiliation{Department of Physics, Boston University, Boston, MA 02215, USA}}
\newcommand{\bnl}{\affiliation{Physics Department, Brookhaven National Laboratory, Upton, NY 11973, USA}}
\newcommand{\uci}{\affiliation{Department of Physics and Astronomy, University of California, Irvine, Irvine, CA 92697-4575, USA}}
\newcommand{\csu}{\affiliation{Department of Physics, California State University, Dominguez Hills, Carson, CA 90747, USA}}
\newcommand{\cnu}{\affiliation{Department of Physics, Chonnam National University, Kwangju 500-757, Korea}}
\newcommand{\gmu}{\affiliation{Department of Physics, George Mason University, Fairfax, VA 22030, USA}}
\newcommand{\gifu}{\affiliation{Department of Physics, Gifu University, Gifu, Gifu 501-1193, Japan}}
\newcommand{\uh}{\affiliation{Department of Physics and Astronomy, University of Hawaii, Honolulu, HI 96822, USA}}
\newcommand{\ui}{\affiliation{Department of Physics, Indiana University, Bloomington,  IN 47405-7105, USA} }
\newcommand{\kek}{\affiliation{High Energy Accelerator Research Organization (KEK), Tsukuba, Ibaraki 305-0801, Japan}}
\newcommand{\kobe}{\affiliation{Department of Physics, Kobe University, Kobe, Hyogo 657-8501, Japan}}
\newcommand{\kyoto}{\affiliation{Department of Physics, Kyoto University, Kyoto 606-8502, Japan}}
\newcommand{\lanl}{\affiliation{Physics Division, P-23, Los Alamos National Laboratory, Los Alamos, NM 87544, USA}}
\newcommand{\lsu}{\affiliation{Department of Physics and Astronomy, Louisiana State University, Baton Rouge, LA 70803, USA}}
\newcommand{\umd}{\affiliation{Department of Physics, University of Maryland, College Park, MD 20742, USA}}
\newcommand{\MIT}{\affiliation{Department of Physics, Massachusetts Institute of Technology, Cambridge, MA 02139, USA}}
\newcommand{\duluth}{\affiliation{Department of Physics, University of Minnesota, Duluth, MN 55812-2496, USA}}
\newcommand{\miyagi}{\affiliation{Department of Physics, Miyagi University of Education, Sendai,Miyagi 980-0845, Japan}}
\newcommand{\suny}{\affiliation{Department of Physics and Astronomy, State University of New York, Stony Brook, NY 11794-3800, USA}}
\newcommand{\nagoya}{\affiliation{Department of Physics, Nagoya University, Nagoya, Aichi 464-8602, Japan}}
\newcommand{\niigata}{\affiliation{Department of Physics, Niigata University, Niigata, Niigata 950-2181, Japan}}
\newcommand{\osaka}{\affiliation{Department of Physics, Osaka University, Toyonaka, Osaka 560-0043, Japan}}
\newcommand{\seoul}{\affiliation{Department of Physics, Seoul National University, Seoul 151-742, Korea}}
\newcommand{\shizuokaseika}{\affiliation{International and Cultural Studies, Shizuoka Seika College, Yaizu, Shizuoka 425-8611, Japan}}
\newcommand{\shizuoka}{\affiliation{Department of Systems Engineering, Shizuoka University, Hamamatsu, Shizuoka 432-8561, Japan}}
\newcommand{\skku}{\affiliation{Department of Physics, Sungkyunkwan University, Suwon 440-746, Korea}}
\newcommand{\tohoku}{\affiliation{Research Center for Neutrino Science, Tohoku University, Sendai, Miyagi 980-8578, Japan}}
\newcommand{\tokyo}{\affiliation{University of Tokyo, Tokyo 113-0033, Japan}}
\newcommand{\tokai}{\affiliation{Department of Physics, Tokai University, Hiratsuka, Kanagawa 259-1292, Japan}}
\newcommand{\tit}{\affiliation{Department of Physics, Tokyo Institute for Technology, Meguro, Tokyo 152-8551, Japan}}
\newcommand{\warsaw}{\affiliation{Institute of Experimental Physics, Warsaw University, 00-681 Warsaw, Poland}}
\newcommand{\uw}{\affiliation{Department of Physics, University of Washington, Seattle, WA 98195-1560, USA}}
\newcommand{\ferminow}{\altaffiliation{ Present address: Enrico Fermi Institute, University of Chicago, Chicago, IL 60637, USA}}
%
\author{Y.Ashie}\icrr
\author{J.Hosaka}\icrr
\author{K.Ishihara}\icrr
\author{Y.Itow}\icrr
\author{J.Kameda}\icrr
\author{Y.Koshio}\icrr
\author{A.Minamino}\icrr
\author{C.Mitsuda}\icrr
\author{M.Miura}\icrr
\author{S.Moriyama}\icrr
% ...
\collaboration{The Super-Kamiokande Collaboration}\noaffiliation

\begin{abstract}
Muon neutrino disappearance probability as a function of 
neutrino flight length $L$ over neutrino energy $E$ was
studied. A dip in the $L/E$ distribution was observed
 in the data, as predicted from the sinusoidal flavor 
transition probability of neutrino oscillation. 
 The observed $L/E$ distribution constrained 
 $\nu_\mu \leftrightarrow \nu_\tau$
 neutrino oscillation parameters;
 $1.9\times 10^{-3} < \Delta m^2 < 3.0\times 10^{-3}{\rm eV}^2$
¡¡and $\sin^22\theta > 0.90$ at 90\% confidence level.
\end{abstract}

\maketitle

Recent neutrino experiments using 
atmospheric~\cite{Fukuda:1998mi, Fukuda:1994mc, Becker-Szendy:1992hq,
Ambrosio:1998wu, Sanchez:2003rb},
solar~\cite{Cleveland:1998nv,Fukuda:1996sz, Hampel:1998xg, 
Abdurashitov:2002nt,Altmann:2000ft, Smy:2003jf, Ahmed:2003kj},
reactor~\cite{Eguchi:2002dm},
and accelerator neutrinos~\cite{Ahn:2002up},
have demonstrated that neutrinos change flavor as they travel from 
the source to the detector, a phenomenon consistent with the 
hypothesis of neutrino oscillation. 
Neutrino oscillation is a natural consequence of neutrinos that have finite 
mass and flavor eigenstates that are superpositions of the mass eigenstates. 
The phenomenon is referred to as oscillation because the
survival probability of a given flavor, such as $\nu_\mu$, is given by:
\begin{equation}\label{eq:oscillation}
 P(\nu_{\mu} \rightarrow \nu_{\mu})
= 1-\sin^22\theta\sin^2\left(\frac{1.27\Delta{m}^2({\rm eV}^2) L({\rm km})
}{E({\rm GeV})}\right),
\end{equation}
where $E$ is the neutrino energy, $L$ is the travel distance, 
$\Delta m^2$ is the
difference of the squared mass eigenvalues, and $\theta$ is the mixing angle
between flavor and mass states. This equation is true
in vacuum for all cases, is true in matter for $\nu_\mu \leftrightarrow
\nu_\tau$, but may be modified for oscillation involving $\nu_e$ which travel 
through matter.

However, the sinusoidal $L/E$ dependence of the survival probability has not
yet been observed. 
For solar neutrinos, the survival probability 
 is non-sinusoidal
as the two eigenstates in matter are no longer coherent after many 
oscillation cycles~\cite{deHolanda:2003nj}.
Reactor and accelerator
neutrino experiments have insufficient statistics at this time. The standard
analysis~\cite{Fukuda:1998mi, Fukuda:2004xx}  
 of the large sample of atmospheric neutrinos recorded by the
Super-Kamiokande experiment has not been optimized to resolve the effect,
although the zenith angle dependence strongly indicates maximal $\nu_\mu
\leftrightarrow \nu_\tau$ mixing with $\Delta m^2$ in the vicinity of $2$ to
$2.5 \times 10^{-3}$ eV$^2$. The analysis described herein used a selected
sample of these atmospheric neutrino events, those with good resolution in
$L/E$, to search for the dip in oscillation probability expected when the
argument of the second sine-squared term in Eq.~\ref{eq:oscillation} is $\pi/2$.

Super-Kamiokande (Super-K) 
is a cylindrical 50\,kton water Cherenkov detector located
 at a depth of 2700\,m water equivalent. The water tank is optically
 separated into two concentric cylindrical detector regions. 
 The inner detector (ID)
 is instrumented with 11,146 inward facing 20\,inch diameter photomultiplier
 tubes (PMT). 
 The outer detector (OD) is instrumented with 1,885 outward 
 facing 8\,inch PMTs.
 
 In the present analysis, 1489\,live-day exposure of fully contained (FC)
 $\mu$-like and partially contained (PC) atmospheric neutrino data were used.
 FC events deposit all of their Cherenkov light inside the ID, while PC
 events have an exiting particle that deposits visible energy in the OD. 
 The
 direction and the momentum of charged particles were reconstructed from the
 Cherenkov ring image. Each observed ring was identified as either $e$-like or
 $\mu$-like based on the shape of the ring pattern. For FC multi-ring
 events, the particle type of the most energetic ring was used to identify
 $\mu$-like events. Since more than 97\,\% of PC events were estimated to be
 $\nu_{\mu}$ charged current (CC) interactions, all PC events were classified
 as $\mu$-like.  
 The atmospheric neutrino prediction in Super-K is
 modeled using a Monte Carlo (MC) simulation~\cite{Fukuda:2004xx}.

\begin{figure}[htb]
  \includegraphics[width=2.65in]{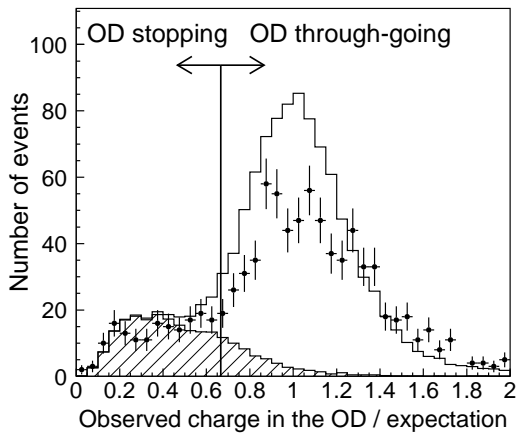} 
  \caption{Observed charge (photoelectrons) 
 in the OD divided by the expectation from
 the projected track length in the OD for the data (points),
 the OD through-going MC events (white region in histogram)
 and the OD stopping MC events (hatched region). The MC
 does not include oscillations and is normalized by the 
 live-time. }
 \label{fig:pc_stop_through}
\end{figure}

Event selection and classification in the present analysis
are slightly different from those in the Super-K 
standard oscillation analysis using zenith
 angle distributions. The fiducial volume for the FC events was expanded
 from 22.5\,kton to 26.4\,kton
 (event vertex should be more than 1.5\,m from the top and
 bottom walls of the ID and 1\,m from the side wall)
 in order to increase the statistics of the data,
 especially of high energy muons.
 Estimated non-neutrino background in the
 expanded fiducial volume was negligibly small, less than 0.1\%.
 PC events were classified into two categories using the OD charge 
(photoelectron) information:
 ``OD stopping events'' and ``OD through-going events''.
 Muons in the ``OD stopping events''
 have stopped in the OD, while muons in the
 ``OD through-going events'' have passed through the OD.
 Figure~\ref{fig:pc_stop_through} shows
  the ratio of the observed charge in the OD to the expectation from
 the projected track length in the OD.
Events with lower charge than the
 criterion were classified as ``OD stopping events''.
 Since these two samples have different resolution in $L/E$,
 different cuts were applied for each sample,
 improving the overall efficiency.

\begin{figure}[htb]
  \includegraphics[width=3.3in]{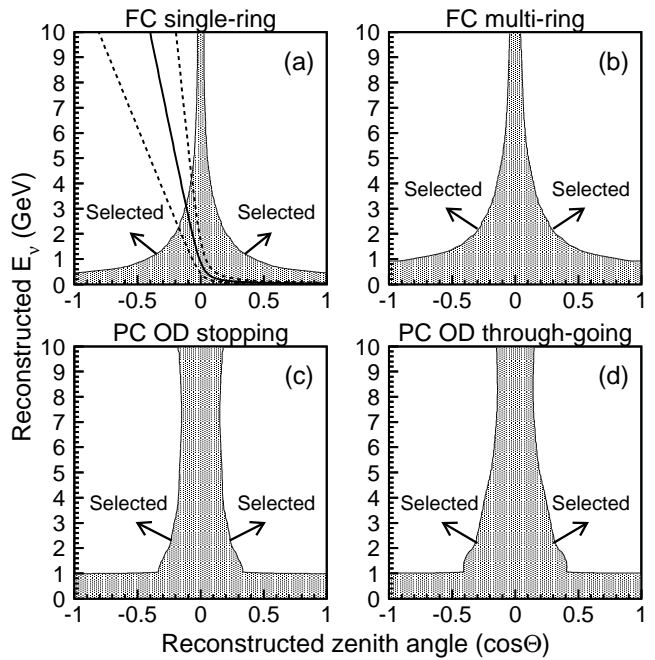} 
  \caption{Contour plots of 70\,\% $L/E$ resolution in the
 ($\cos\Theta,\,\,E_{\nu}$) plane for (a) FC single-ring,
  (b) FC multi-ring, (c) PC OD stopping
 and (d) PC OD through-going samples. Three additional lines in (a)
 show the survival probabilities of muon neutrinos predicted
 from neutrino oscillation with ($\sin^2 2\theta,\,\Delta m^2) = (1.00,\,2.4\times10^{-3} $\,eV$^2$). Full and half oscillation 
 occur on the solid and dashed lines, respectively. }
 \label{fig:resolution_le}
\end{figure}

The neutrino energy was estimated from the total energy of charged
 particles observed in the ID. The energy deposited in the OD was
 taken into account for PC events. The projected track length in the OD
 was used to estimate the energy deposited in the OD.
 The relationship between the neutrino
 energy and the observed energy was determined based on
 the MC simulation.  
 The flight length of neutrinos, which ranges from approximately 15\,km
 to 13,000\,km depending on the zenith angle, was estimated
 from the reconstructed neutrino direction. 
 The neutrino direction was taken to be along
 the total momentum vector from all observed particles.
The resolution of the reconstructed $L/E$ was calculated at each point in the
($\cos\Theta, E$) plane, where $\Theta$ is the zenith angle.  
The energy resolution
becomes poorer for higher energy PC events, due mainly to the saturation in
the electronics that records the PMT charge. 
Therefore, extremely high energy events 
(observed energy  $> 50$\,GeV)
were excluded in this analysis.
All the FC $\mu$-like events have observed energy less than
25~GeV, so this cut is only relevant to PC events.
Figure~\ref{fig:resolution_le} shows 70\,\% $L/E$ resolution contours, as
used for the selection criteria in this analysis. The reasons for the poor
$L/E$ resolutions are either large $dL/d \Theta $ for horizontal-going
events or large scattering angles for low energy events. The bold solid 
central
line in Fig.~\ref{fig:resolution_le}a indicates the minimum survival
probability of muon neutrinos predicted from neutrino oscillations with
$\Delta m^2 = 2.4 \times 10^{-3}$ eV$^2$. It is clear that detecting high
energy muon events is crucial to observe the first maximum oscillation in
$L/E$. The resolution cut of $\Delta(L/E) < 70$\,\% was determined
from the MC simulation to maximize the sensitivity to distinguish
neutrino oscillation from other hypotheses.

\begin{figure}
\begin{center}
  \includegraphics[width=3.0in]{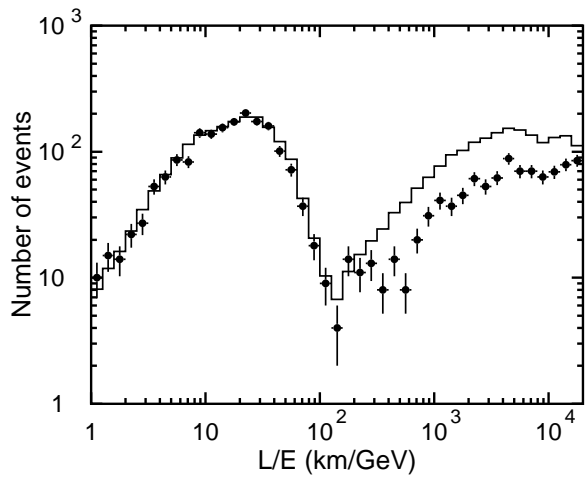}
  \caption{Number of events as a function of $L/E$ for the data (points)
 and the atmospheric neutrino MC events without oscillations
 (histogram). The MC is normalized by the detector live-time.}
   \label{fig:le_dist}
\end{center}
\end{figure}

\begin{figure}
\begin{center}
  \includegraphics[width=3.0in]{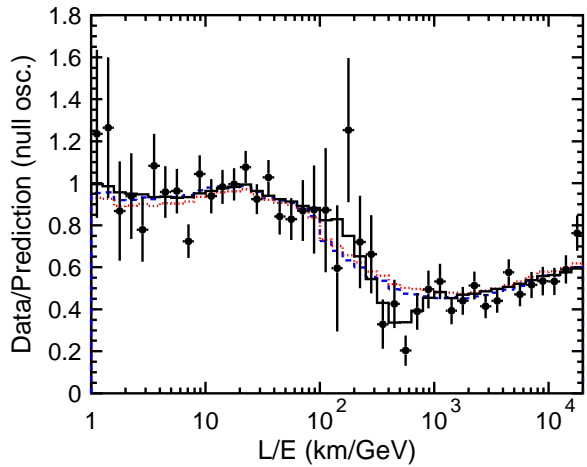}
  \caption{Ratio of the data to the MC events without
 neutrino oscillation (points) as a function of the
 reconstructed $L/E$ together with the best-fit expectation
 for 2-flavor $\nu_\mu \leftrightarrow \nu_\tau$
 oscillations (solid line). 
 The error bars are statistical only.
 Also shown are the best-fit expectation for
 neutrino decay (dashed line) and neutrino decoherence (dotted line).
}
   \label{fig:bestfit_models}
\end{center}
\end{figure}

Table~\ref{table:evsummary_le} summarizes the number of events used in
this analysis after the $L/E$ resolution cut.
\begin{table}
\begin{center}
\begin{tabular}{llccc}
\hline
\hline
      &  & ~Data~ & ~MC~ & ~$\nu_{\mu}+\overline{\nu}_{\mu}$ CC~ \\
\hline
FC & ~~single-ring $\mu$-like &  1619  & 2105.8 & (98.3\,\%)\\
   & ~~multi-ring $\mu$-like & 502 & 813.0 & (94.2\,\%)\\
\hline
PC & ~~OD stopping      & 114 & 137.0 & (95.4\,\%)\\
   & ~~OD through-going~~ & 491 & 670.4 & (99.1\,\%)\\
\hline
\hline
\end{tabular}
\caption{Summary of atmospheric neutrino events used in the present analysis.
 Only $\mu$-like events are used. Numbers of the MC events are
 normalized by the live-time.
 Neutrino oscillation is not included in the MC.
 Numbers in the parentheses show the estimated fraction of $\nu_{\mu}+\overline{\nu}_{\mu}$
 CC interactions in each sample.}
\label{table:evsummary_le}
\end{center}
\end{table}
Figure~\ref{fig:le_dist} shows the number of events as a function
 of $L/E$ for the data and MC predictions. Two clusters of
events are visible below and above 150~km/GeV. They mostly 
correspond to downward-going and upward-going events, respectively.
Below 150~km/GeV, the data and MC agree well.
In Fig.~\ref{fig:bestfit_models} 
the data over non-oscillated MC ratio 
as a function of $L/E$ is plotted together
 with the best-fit expectation for 2-flavor $\nu_\mu \leftrightarrow \nu_\tau$
 oscillations with systematic errors.
 A dip, which should correspond to the first maximum oscillation,  
 is observed around $L/E$~=~500~km/GeV. 
We note that the position of the dip is
about a factor of 3 to 4 away from that of the predicted event number
minimum as seen in Fig.~\ref{fig:le_dist}. 
Due to the $L/E$ resolution 
 of the detector, the second and higher maximum oscillation points 
 should not be observable in this experiment.

In order to confirm that the observed dip was not due to 
systematic effects, several tests were carried out.
Several $L/E$ distributions were made by changing the 
$L/E$ resolution cut value. Plots based on the resolution cuts 
at 60, 80 and 90\% showed consistent dip structures as that 
based on the 70\% cut. 
Also, $L/E$ plots based on several other $L/E$ bin sizes gave
essentially the same results.  
In addition, the sign of the direction vector for
each event was changed artificially. In this artificial data
sample, the ``upward-going'' events should have little
disappearance effect and therefore the $L/E$ distribution 
 should not show any dip structure
around $L/E$ = 500~km/GeV. 
The $L/E$ distribution for this
artificial data sample did not show any significant dip structure
around 500~km/GeV. Finally, the $L/E$ plot was made using FC single-ring
$e$-like events. The $e$-like distribution was consistent with
flat over the whole $L/E$ range.  
Thus we are confident that the observed
dip is not due to systematic effects in the event selection.   

The data/prediction at large $L/E$ in Fig.~\ref{fig:bestfit_models} 
shows a slight
rise from the expected flat distribution. We have studied possible
causes of this deviation, and concluded that an energy-dependent systematic
effects, such as the predicted neutrino interaction cross section,
are the main
sources of the non-flatness. The best-fit $L/E$ distribution for
oscillations, allowing systematic terms to vary 
within the estimated uncertainty (as described below), 
also shows this rise with respect to no-oscillation prediction, as seen in
the curves overlaid in Fig.~\ref{fig:bestfit_models}. 
The rise at large $L/E$ is consistent with the data.

 The observed $L/E$ distribution was fit assuming 
 $\nu_\mu \leftrightarrow \nu_\tau $
 oscillations. The $L/E$ distribution was divided into
 43 bins from $log(L/E)=0.0$ to 4.3\,.
 The likelihood of the fit and the $\chi^2$ were defined as:
\begin{eqnarray}
{\cal L}(N^{\rm prd},N^{\rm obs}) = \prod_{i=1}^{43}\frac{\exp{(-N_{i}^{\rm prd})(N_{i}^{\rm prd})^{N_{i}^{\rm obs}}}}{N_{i}^{\rm obs}!} \nonumber\\
\times\prod_{j=1}^{24}\exp{\left( - \frac{\epsilon_j^2}{2\sigma_j^2} \right)}, \\
 N_{i}^{\rm prd} = N_{i}^{\rm 0} \cdot
   P(\nu_\mu \rightarrow \nu_\mu) \cdot 
   (1+\sum_{j=1}^{25}f_{j}^{i}\cdot\epsilon_{j} ),~~~ \\
\chi^2 \equiv -2\ln\left(\frac{{\cal L}(N^{\rm prd},N^{\rm obs})}{{\cal L}(N^{\rm obs},N^{\rm obs})}\right),~~~~~~~~~~
\label{equation:chi2def_le}
\end{eqnarray}
where $N^{\rm obs}_i$ is the number of the observed events in
 the $i$-th bin and $N^{\rm prd}_i$ is the number of predicted 
 events, in which neutrino oscillation and systematic
 uncertainties are considered.  $N^{\rm 0}_i$ is the MC 
 predicted number of events without oscillation for the $i$-th bin. 
 Various systematic uncertainties are represented by 25 parameters
 $\epsilon_j$, which include 7 uncertainty parameters from the
flux calculation (among these, absolute normalization is treated as a
free parameter), 3 from the detector calibration and background, 
2 from the data reduction, 5 from the event reconstruction, and 8
from the neutrino interaction simulation. 
A more detailed description
of the systematic error terms can be found in Ref.~\cite{Fukuda:2004xx}.
The second term in the likelihood definition represents
the contributions from the systematic errors, where $\sigma_j$
is the estimated uncertainty in the parameter $\epsilon_j$.
The fractional effect
of systematic error term $\epsilon_j$ on the $i$-th bin is given
by $f_j^i$.

\begin{figure}
\begin{center}
  \includegraphics[width=2.65in]{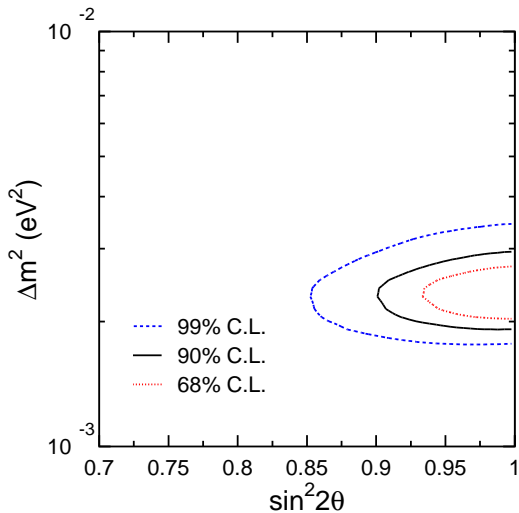}
  \caption{68, 90 and 99\%~C.L. allowed oscillation parameter regions for
 2-flavor $\nu_\mu \leftrightarrow \nu_\tau$ oscillations obtained
 by the present analysis.}
   \label{fig:osc_allowed_le}
\end{center}
\end{figure}

A scan was carried out on a $(\sin^22\theta,\,\log \Delta m^2)$ grid,
 minimizing $\chi^2$ 
 by optimizing the systematic error parameters at each point.
 The minimum $\chi^2$ was $37.9/40$\,DOF at
 $(\sin^{2}2\theta,\,\Delta m^2)=(1.00,\,2.4\times10^{-3} $\,eV$^2$).
 Including unphysical parameter region ($\sin^22\theta > 1$),
 the best-fit was obtained at
 $(\sin^{2}2\theta,\,\Delta m^2)=(1.02,\,2.4\times10^{-3} $\,eV$^2$),
 in which the minimum $\chi^2$ was 0.12 lower than that in the physical region.
 Figure~\ref{fig:osc_allowed_le} shows the contour plot of the
 allowed oscillation parameter regions. Three contours correspond to the
 68\,\%, 90\,\% and 99\,\% confidence level (C.L.) allowed regions, which are
 defined to be $\chi^2 = \chi^2_{min} +$ 2.48, 4.83, and 9.43, respectively,
 where $\chi^2_{min}$ is the minimum $\chi^2$ in the physical region.
 These intervals are derived based on a two dimensional extension of
 the method described in Ref.~\cite{Barnett:1996hr}. 
The 90\,\% C.L. allowed parameter
 region was obtained as 
 $1.9 \times 10^{-3}\,{\rm eV}^2 < \Delta m^2 
< 3.0 \times 10^{-3}\,{\rm eV}^2 $ and 
$\sin^{2}2\theta~ > 0.90$.
The result is consistent with that of the oscillation analysis
 using zenith angle distributions~\cite{Fukuda:1998mi, Fukuda:2004xx}.

In order to test the significance of the dip in $L/E$, a null 
hypothesis that includes the basic shape of the $L/E$ distribution 
is needed. The no-oscillation case was very strongly disfavored 
by the data at large $L/E$. We used two alternative hypotheses that 
basically reproduce the zenith angle dependent deficit,
predicting that about half of muon neutrinos smoothly disappear 
at large $L/E$. These hypotheses are neutrino
decay~\cite{Barger:1998xk, Barger:1999bg} and neutrino 
decoherence~\cite{Grossman:1998jq, Lisi:2000zt}.
The $\nu_\mu$ survival probability for neutrino decay 
is expressed as  $P(\nu_{\mu} \rightarrow \nu_{\mu})$ 
= 
$\left[\sin^{2}\theta+\cos^{2}\theta \exp\left(-m/2\tau \cdot L/E\right)\right]^2$ 
where $\tau$ is the lifetime of a neutrino mass state. The 
neutrino decoherence survival probability is
$P(\nu_{\mu} \rightarrow \nu_{\mu})$ 
= 
$1-\frac{1}{2}\sin^{2}2\theta\left[1-\exp\left(-\gamma_{0} L/E\right)\right]$, 
where $\gamma_0$ is the decoherence parameter. 
Figure~\ref{fig:bestfit_models} includes the 
$L/E$ distribution for the best-fit expectation for neutrino decay 
and neutrino decoherence.
The $\chi^2_{min}$ values were $49.1/40$\,DOF
 at $(\cos^{2}\theta,\,m/\tau)=(0.33,\,1.26 \times 10^{-2}$\,GeV/km)
 for neutrino decay and $52.4/40$\,DOF
 at $(\sin^{2}2\theta,\,\gamma_{0})=(1.00,\,1.23\times10^{-21}$\,GeV/km)
 for neutrino decoherence. These $\chi^2_{min}$ values were 11.3 (3.4 standard 
deviations)
 and 14.5 (3.8 standard deviations) larger than $\chi^2_{min}$ for neutrino
 oscillation. 
 In order to check the statistical significance against the
alternative models, 10,000 MC $L/E$ distributions were 
produced assuming neutrino decay with the best fit decay parameters.
Each $L/E$ distribution was fitted assuming neutrino decay and 
oscillation, and the $\chi^2$ difference for these two assumptions
was calculated. For example, only 11 among 10,000 samples had $\chi^2$ for 
neutrino decay smaller by 11.3 or more than the same sample evaluated 
for neutrino oscillation. Therefore, 
the probability that neutrino decay could
mimic neutrino oscillations is
 approximately 0.1\% as naively expected by 3.4 standard deviations.
The neutrino decoherence model is disfavored more strongly.

In summary, we have studied the survival probability of muon 
neutrinos as a function of $L/E$ using atmospheric
neutrino events observed in Super-Kamiokande. A dip in the
$L/E$ distribution was observed around $L/E$ = 500 km/GeV.  
This strongly constrains $\Delta m^2$.
Alternative models that could explain the zenith angle and
energy dependent deficit of the atmospheric muon neutrinos are 
disfavored, since
they do not predict any dip in the $L/E$ distribution. 
We conclude that 
the observed $L/E$ distribution gives the 
first direct evidence that the neutrino survival
probability obeys the sinusoidal function as predicted
 by neutrino flavor oscillations. 

We gratefully acknowledge the cooperation of the Kamioka Mining and
Smelting Company.  The Super-Kamiokande experiment has been built and
operated from funding by the Japanese Ministry of Education, Culture,
Sports, Science and Technology, the United States Department of Energy,
and the U.S. National Science Foundation.

\end{document}